# Requirements for DNA-bridging proteins to act as topological barriers of the bacterial genome

Marc Joyeux[(#)]
*Laboratoire Interdisciplinaire de Physique,*
*CNRS and Université Grenoble Alpes, Grenoble, France*

Ivan Junier
*TIMC-IMAG*
*CNRS and Université Grenoble Alpes, Grenoble, France*

**Running title**: Topological barriers on bacterial DNA.

**Abstract:** Bacterial genomes have been shown to be partitioned into several kilobases long chromosomal domains that are topologically independent from each other, meaning that change of DNA superhelicity in one domain does not propagate to neighbors. Both *in vivo* and *in vitro* experiments have been performed to question the nature of the topological barriers at play, leading to several predictions on possible molecular actors. Here, we address the question of topological barriers using polymer models of supercoiled DNA chains that are constrained such as to mimic the action of predicted molecular actors. More specifically, we determine under which conditions DNA-bridging proteins may act as topological barriers. To this end, we developed a coarse-grained bead-and-spring model and investigated its properties through Brownian dynamics simulations. As a result, we find that DNA-bridging proteins must exert rather strong constraints on their binding sites: they must block the diffusion of the excess of twist through the two binding sites on the DNA molecule and, simultaneously, prevent the rotation of one DNA segment relative to the other one. Importantly, not all DNA-bridging proteins satisfy this second condition. For example, single bridges formed by proteins that bind DNA non-specifically, like H-NS dimers, are expected to fail with this respect. Our findings might also explain, in the case of specific DNA-bridging proteins like LacI, why multiple bridges are required to create stable independent topological domains. Strikingly, when the relative rotation of the DNA segments is not prevented, relaxation results in complex intrication of the two domains. Moreover, while the value of the torsional stress in each domain may vary, their differential is preserved. Our work also predicts that nucleoid associated proteins known to wrap DNA must form higher protein-DNA complexes to efficiently work as topological barriers.

[(#)] marc.joyeux@univ-grenoble-alpes.fr

**Statement of significance:** Although the question of independent topological domains in bacterial genomes has been addressed for nearly two decades, the nature of topological barriers is still elusive. Most plausible models for the formation of topological barriers are actively transcribing RNAP and the formation of DNA loops by certain nucleoid proteins. In the present paper, we focus on this latter mechanism and establish under which conditions DNA-bridging proteins may serve as topological barriers. We show that, contrary to popular belief, formation of a loop is not sufficient and that proteins must additionally block the diffusion of twist and prevent the relative rotation of DNA segments. These findings set strong theoretical limits on the ability of bridging proteins to create topologically independent genomic domains.

# INTRODUCTION

The genetic information of most bacteria is encoded in a circular DNA molecule comprising up to several millions of base pairs (bp). Such closed molecules are topologically constrained, because they lack free ends capable of rotating and releasing the torsional stress (1). *In vivo* alterations of the torsional state of circular DNA is mediated by the recruitment of enzymes called topoisomerases, which can either increase or decrease the twist of the double helix by opening transiently one or two strands (1). The net result of the action of all types of topoisomerases is that the genomic DNA of most bacteria is significantly undertwisted (negatively supercoiled) and winds about itself to transfer part of the torsional stress to the bending degrees of freedom, thereby forming plectonemes (1). If the DNA were not subject to any additional constraint beyond closure, then one single nick of one strand or one single break of the two strands would suffice to release the torsional stress of the full molecule. It has however long be known that at least several tens of nicks are required to achieve this goal in *Escherichia coli* (2,3). This indicates that there exist barriers which block the diffusion of the torsional stress along the DNA molecule and organize the chromosome of *E. coli* into many independent topological domains, whose torsional state is not affected by the relaxation of other domains. It is currently estimated that the genomic DNA of bacteria like *E. coli* is composed of several hundreds of different topological domains with variable size (average size ≈10 kbp) and position (4). More generally, the partitioning of the bacterial chromosome into ≈10 kbp independent topological domains is believed to hold in most bacteria (5) and to be related, in part, to the insulation of fundamental co-expression units that are neighbors along the genome (6).

The work reported here deals with the topological barriers which make this partitioning possible, that is, the barriers which block the diffusion of torsional stress along the DNA molecule and are responsible for its division into independent topological domains. Although this question has now been addressed for nearly two decades (7-9), the mechanisms underlying the formation of topological barriers remain mostly elusive. Most plausible models for the formation of topological barriers are (i) actively transcribing RNAP, which generate both positive and negative supercoils (10) and may consequently block the displacement and dissipation of plectonemes (11,12), and (ii) formation of DNA loops by certain nucleoid proteins, which may serve as topological barriers (13-16). In the present paper, we focus on this latter mechanism and establish under which conditions DNA-bridging proteins may serve as topological barriers.

Several DNA-bridging proteins have been studied in some detail, including transcription regulators like the LacI repressor (17), H-NS (18) and H-NS-like proteins (19), Lsr2 (20), Fis (21) and Lrp (22). In their functional form, all of these proteins have at least two independent DNA-binding domains, so that they can interact with two DNA duplexes simultaneously and form a bridge between two sites that are widely separated from the genomic point of view. Most of these proteins also bind non-specifically but with high affinity to the DNA molecule. However, among all these DNA-bridging proteins, LacI is to date the only one that has been proved to work as a topological barrier (15,23,24).

The guiding line of the present paper is consequently the experimental demonstration in (15) that *the binding of a DNA-binding protein to its recognition sites in two different locations on a supercoiled DNA molecule can confine free supercoils to a defined region and divide the DNA molecule into two distinct topological domains* (15). In order to gain more detailed information on this mechanism, we determined the minimal physical properties that must be added to a standard model of circular DNA to reproduce the results described in (15). The starting model for this study was similar to those we proposed recently to investigate facilitated diffusion (25-27), the interactions of DNA and H-NS nucleoid proteins (28-30), the formation of the bacterial nucleoid (31-35), and the interplay of DNA demixing and supercoiling in compacting the nucleoid (36).

The results presented in this article reveal that the formation of DNA loops is by no means sufficient to create topologically independent domains and that DNA-bridging proteins must exert rather strong constraints on their binding sites in order to act as topological barriers: They must block the diffusion of the excess of twist through both binding sites and must additionally block the rotation of one DNA segment relative to the other one.

**METHODS**

The coarse-grained bead-and-spring model developed for the present study is described in detail in Model and Simulations in the Supporting Material. In brief, bacterial DNA is modeled, as in (30,36), as circular chains of *n* beads with radius $a = 1.0$ nm separated at equilibrium by a distance $l_0 = 2.5$ nm, where two beads represent 15 DNA bp. Simulations were performed with unconstrained circular chains of length $n = 600$, equivalent to 4500 bp, which represent the plasmids used in (15). DNA beads interact through stretching, bending, torsional, and electrostatic terms. The bending rigidity constant ( $g = 20\, k_B T$ ) was chosen so that the model reproduces the known persistence length of double-stranded DNA ($\xi = 50$ nm).

As in (31,36), the torsional energy term was borrowed from (37) and requires the introduction of a body-fixed frame $(\mathbf{u}_k, \mathbf{f}_k, \mathbf{v}_k)$, where $\mathbf{u}_k$ denotes the unit vector pointing from bead $k$ to bead $k+1$. The torsional rigidity opposing rotation of $(\mathbf{u}_k, \mathbf{f}_k, \mathbf{v}_k)$ around $\mathbf{u}_k$ ($\tau = 25\,k_B T$) was adjusted so that at equilibrium the writhe contribution accounts for approximately 70% of the linking number difference (38), as is illustrated in Fig. S1. The values of the bending and torsional rigidities are close together, in agreement with experimental results (38). It may be worth emphasizing that introduction of the body-fixed frame $(\mathbf{u}_k, \mathbf{f}_k, \mathbf{v}_k)$ is crucial for modeling correctly the torsion of double-stranded DNA with a "single-stranded" bead-and-spring model. This procedure is quite realistic and has already proved very useful for studying, for example, the interplay of DNA demixing and supercoiling in nucleoid compaction (36), the buckling transition in double-stranded DNA and RNA (39), the influence of nucleoid-associated proteins on DNA supercoiling (40), DNA supercoil relaxation (41), the competition between B-Z and B-L transitions in a single DNA molecule (42), the relaxation of DNA supercoils by topoisomerase IB (43), the sequence-dependent duplex denaturation in superhelical DNA (44), transcription-driven twin supercoiling of DNA loops (45), site juxtaposition in supercoiled DNA (46), and salt effects on the structure and internal dynamics of superhelical DNA (47).

Electrostatic repulsion between DNA beads is written as a sum of Debye-Hückel terms, which depend on effective electrostatic charges placed at the center of each bead. The values of these charges ($q \approx -3.52\,\bar{e}$, where $\bar{e}$ is the absolute charge of the electron) were derived from the net linear charge density along a DNA molecule immersed in a buffer with monovalent cations according to Manning's counterion condensation theory (48,49). The value of the Debye length ($r_D = 1.07$ nm) corresponds to a concentration of monovalent salt of 100 mM, which is the value that is generally assumed for the cytoplasm of bacterial cells.

The torsional state of the DNA chain is characterized by the value of the superhelical density $\sigma = \Delta Lk / Lk_0$, where the linking number $Lk_0 = 7.5n/10.5 \approx 429$ is the ratio of the number of base pairs of the DNA chain and the mean number of base pairs per turn of the torsionally relaxed double helix, and the linking number difference $\Delta Lk = \Delta Tw + Wr$ is the sum of the excess of twist ($\Delta Tw$) and the writhe ($Wr$), which quantifies the winding of the DNA axis around itself. $\Delta Lk$ and $\sigma$ are constants of the motion, while $\Delta Tw$ and $Wr$ fluctuate under the influence of thermal noise and external constraints. Bacterial DNA being negatively supercoiled (undertwisted), $\Delta Lk$ and $\sigma$ are negative in this work. Investigated values of $\sigma$ range from -0.033 to -0.121, that is approximately from the effective value for protein-bound DNA in living *E. coli* cells ($\sigma \approx -0.025$ (50)) to about twice the average value for naked DNA *in vitro* ($\sigma \approx -0.06$ (51)). $\sigma \approx -0.12$ is the maximum value generated by DNA gyrase and RNA polymerase (52-54) and corresponds to a range of superhelicity values where hyperplectonemes, a higher order DNA organization, have been observed (55,56). The properties of the model were determined by integrating numerically overdamped Langevin equations with time steps $\Delta t = 10$ ps. Temperature *T* was assumed to be 298 K throughout the study.

**RESULTS AND DISCUSSION**

As a starting point, we recall the experimental results in (15), where the torsional relaxation properties of 4100 bp plasmids with initial superhelical density $\sigma \approx -0.06$ ($\Delta Lk \approx -24$) were thoroughly investigated. Of special interest to us were the experiments performed with plasmids containing tandem copies of the binding site of the sequence-specific DNA-binding protein LacI at two different locations, such that the plasmid was divided into two stable loops of respective length 2900 bp and 1200 bp upon binding of LacI. The 1200 bp region moreover contained the recognition site for one nicking enzyme (Nt.BbvC1) and the 2900 bp region the recognition site for a second nicking enzyme (Nb.BtsI). In the absence of LacI, addition of Nt.BbvC1 alone or Nb.BtsI alone was sufficient to release the full torsional stress of the plasmid. In contrast, in the presence of LacI, addition of Nt.BbvC1 alone removed only the 7 negative supercoils of the 1200 bp region, while addition of Nb.BtsI alone removed only the 17 negative supercoils of the 2900 bp region (15). Simultaneous addition of both enzymes was required to relax the full plasmid (15). This experiment demonstrates very clearly that pairs of LacI bridges block the diffusion of torsional stress and divide the plasmids into two independent topological domains.

We report below on our efforts to determine the minimal set of mandatory properties that allow molecular bridges to act as topological barriers and block the diffusion of torsional stress in out-of-equilibrium supercoiled chains.

**Preparation of torsionally out-of-equilibrium circular chains.**

Topological barriers are objects that block the diffusion of torsional stress and are able to maintain different torsional conditions on their left and right sides. Characterization of the underlying mechanism requires prior understanding of how torsional stress diffuses in out-of-equilibrium supercoiled chains. To prepare such chains, we first let circular chains with $n = 600$ beads and different values of $\Delta Lk$ equilibrate for about 100 ms. $\Delta Lk$ was set to $\Delta Lk^0 = -14$, $-20$, $-26$, $-34$, $-40$, $-46$, and $-52$, corresponding to $\sigma^0 \approx -0.033$, $-0.047$, $-0.061$, $-0.079$, $-0.093$, $-0.107$, and $-0.121$, respectively. Equilibrated chains were then formally divided into two moieties of equal length on both sides of beads $\alpha$ ($1 \leq \alpha \leq n/2$) and $\beta = \alpha + n/2$, which were chosen so that the distance between their centers is smaller than 10 nm. This step is illustrated in Fig. 1(a), which shows an equilibrated chain with $\Delta Lk^0 = -40$ ($\sigma^0 \approx -0.093$), where beads $\alpha \leq k \leq \beta$ are colored in green and the other ones in red. The torsional stress in segment $\alpha \leq k \leq \beta$ was then artificially released by rotating the internal basis $(\mathbf{u}_k, \mathbf{f}_k, \mathbf{v}_k)$ around $\mathbf{u}_k$ by an angle $2\pi(k-\alpha)\Delta Lk^0/n$ for $\alpha \leq k \leq \beta$, thereby increasing the twist in this segment by an amount equal to $\Delta Lk^0/2$. Finally, the segment $\alpha \leq k \leq \beta$ was equilibrated again for 5 ms, while the other moiety was kept frozen. As illustrated in Fig. 1(b), this led to torsionally out-of-equilibrium circular chains, with one torsionally relaxed moiety, such that $\Delta Lk_1 \approx \Delta Tw_1 \approx Wr_1 \approx 0$, and one torsionally stressed moiety, such that $\Delta Lk_2 \approx \Delta Lk^0/2$, $\Delta Tw_2 \approx 0.3 \Delta Lk_2$, and $Wr_2 \approx 0.7 \Delta Lk_2$. The $\Delta Tw_j$ and $Wr_j$ ($j = 1, 2$) are the partial excess of twist and writhe for each moiety, and the $\Delta Lk_j = \Delta Tw_j + Wr_j$ are the partial linking number differences. These quantities are introduced in more detail in Model and Simulations in the Supporting Material and satisfy

$$\Delta Tw = \Delta Tw_1 + \Delta Tw_2$$
$$Wr = Wr_1 + Wr_2 + Wr_{12} \qquad (1)$$
$$\Delta Lk = \Delta Lk_1 + \Delta Lk_2 + Wr_{12},$$

where $Wr_{12}$ quantifies the winding of one moiety of the chain around the other one, that is, loosely speaking, their intrication. The out-of-equilibrium chains prepared as described above satisfy $Wr_{12} \approx 0$.

The main purpose of this work is to understand how DNA-bridging proteins like LacI (15,23,24) can maintain such an unbalance when the whole chain is relaxed. Since DNA-bridging proteins form DNA loops by dynamically cross-linking widely separated DNA sites, the question amounts here to determine the constraints that must be imposed to beads $\alpha$ and $\beta$ to divide the circular DNA chain into two topologically independent loops. Note that DNA-bridging proteins are not introduced explicitly in the model: Rather, constraints on beads $\alpha$ and $\beta$ are meant to model the mechanisms that allow the proteins to act as topological barriers.

**Twist equilibrates much more rapidly than writhe.**

In order to understand in more detail how the diffusion of torsional stress proceeds, the out-of-equilibrium circular chains prepared as described above were allowed to relax without constraint and all relevant quantities were monitored. From a practical point of view, four different initial out-of-equilibrium conformations were prepared for each value of $\Delta Lk^0$ and 100 different relaxation trajectories were integrated for 1 ms for each initial conformation. The evolution of $\Delta Lk_j$, $\Delta Tw_j$, and $Wr_j$ ($j=1$, 2) was then averaged over these 100 trajectories. In the absence of any constraint on beads $\alpha$ and $\beta$, all conformations obtained at the end of the 1 ms integration time satisfy $\Delta Lk_1 \approx \Delta Lk_2$ and $\Delta Tw_1 \approx \Delta Tw_2$, meaning that torsional stress is distributed homogeneously in the whole chain. Information on the dynamics of the relaxation is provided by the examination of the time evolution of $\Delta Tw_2 - \Delta Tw_1$ and $Wr_2 - Wr_1$. As illustrated in Fig. S2 for an initial conformation with $\Delta Lk^0 = -26$ ($\sigma^0 \approx -0.061$), both quantities decay almost exponentially

$$\Delta Tw_2 - \Delta Tw_1 \approx b \exp\left(-\frac{t}{\tau_{Tw}}\right)$$
$$Wr_2 - Wr_1 \approx A + B \exp\left(-\frac{t}{\tau_{Wr}}\right),$$
(2)

where $\tau_{Tw}$ and $\tau_{Wr}$ are the characteristic time constants for the equilibration of the excess of twist and the writhe, respectively. Let us stress here that we are not aware of any reason why $\Delta Tw_2 - \Delta Tw_1$ and $Wr_2 - Wr_1$ should decay according to an exponential law, that is, according to a process involving a single time scale, although such exponential decays might be rationalized by analyzing the relaxation dynamics of the DNA chain as a twist diffusion process (57). Eq. (2) should thus be considered here essentially as a convenient means to extract and compare characteristic equilibration time constants for the different models investigated in the present paper, that is the circular DNA chain without constraints on beads $\alpha$ and $\beta$ discussed in the present subsection and the three different models of DNA-bridging proteins discussed in the following subsections. The evolution of $\tau_{Tw}$ and $\tau_{Wr}$ as a function of $-\sigma^0$ is shown in Fig. 2. Each point in this figure was obtained from a fit of Eq. (2) against the time evolution of $\Delta Tw_2 - \Delta Tw_1$ and $Wr_2 - Wr_1$ for a different initial conformation. This figure indicates that $\tau_{Tw}$ increases slightly with $-\sigma^0$ but remains comprised between 1 and 2 µs for all investigated values of $-\sigma^0$. Equilibration of the excess of twist between the two moieties of the circular chain is consequently rather fast, the reason being that it involves only some rotation of the chain around its own axis – more precisely, rotation of the internal basis $(\mathbf{u}_k, \mathbf{f}_k, \mathbf{v}_k)$ around $\mathbf{u}_k$ for all $k$. In contrast, Fig. 2 indicates that $\tau_{Wr}$ is typically one to two orders of magnitude larger than $\tau_{Tw}$, which means that equilibration of the writhe is much slower than equilibration of the excess of twist. The reason for this longer time constant is that equilibration of the writhe requires global rearrangements of the whole chain, as can be checked in Fig. 1, instead of it mere rotation around its own axis. Moreover, the larger the difference in torsional stress between the two moieties, the larger the torsional force and torque exerted by each moiety on the other one, and the faster the motion of the beads involved in the rearrangement. This is probably the reason, why $\tau_{Wr}$ decreases significantly with increasing $|\sigma^0|$, from about 250 µs at $\sigma^0 \approx -0.033$ (in the same range as reported in (41)) down to about 40 µs at $\sigma^0 \approx -0.121$. Finally, the values of $\tau_{Wr}$ vary more broadly from one initial conformation to the other one for smaller values of $|\sigma^0|$ than for larger ones, as can be checked in Fig. 2.

At equilibrium, the linking number difference decomposes into approximately 70% of writhe and 30% of excess of twist (see Fig. S1), but equilibration of the excess of twist is much faster than equilibration of the writhe. After a couple of microseconds, the linking number difference in the initially relaxed moiety of the chain is therefore composed of $0.5 \times 0.3 \times 0.5 \times \Delta Lk^0 = 0.075 \, \Delta Lk^0$ excess of twist and no writhe (0% writhe, 100% excess of twist), while in the initially stressed moiety it decomposes into $0.7 \times 0.5 \times \Delta Lk^0 = 0.35 \, \Delta Lk^0$ of writhe and $0.5 \times 0.3 \times 0.5 \times \Delta Lk^0 = 0.075 \, \Delta Lk^0$ excess of twist (82% writhe, 18% excess of twist). As a consequence, during most part of the equilibration process the initially torsionally relaxed moiety of the chain has not enough writhe compared to the excess of twist, while the initially torsionally stressed moiety has too much writhe compared to the excess of twist. The need to equilibrate the writhe/twist ratio inside each moiety of the DNA chain may in turn accelerate the equilibration of writhe in the whole chain by favoring diffusion of the writhe from the initially stressed moiety to the initially relaxed one.

Several comments are in order here. First, the faster equilibration of twist compared to writhe for closed molecules or filaments with both torsional and bending energy (and comparable torsional and bending force constants) is well-known in the field of DNA biophysics. Yet, there are relatively few papers which discuss this point explicitly and provide estimates for the characteristic relaxation time scales. We note, however, that the theoretical study in (58) already pointed out that twist and writhe equilibrate on very different time scales. Second, it is emphasized that time scales deduced from coarse-grained models always underestimate real time scales by two to three orders of magnitudes, because coarse-grained models neglect many details which systematically slow down the dynamics. This point should of course be kept in mind when analyzing the results presented here and comparing them to experimental results. Third, we tested the importance of rotational noise (the last term in the right-hand side of Eq. (S9)) by launching simulations where this term was not taken into account. Characteristic equilibration times obtained as described above are shown in Fig. S3. Comparison of Figs. 2 and S3 indicates that results obtained with and without rotational noise are identical within computational uncertainties. Rotational noise was consequently ignored in all subsequent simulations.

**Closing a DNA loop does NOT suffice to create an independent topological domain.**

Based on recent experimental results (12-16), one may wonder whether DNA loops are *ipso facto* independent topological domains, or whether some additional, more subtle properties of DNA-bridging proteins are required to achieve this goal. We emphasize that the word "loop" is used here according to its biological meaning and not the geometrical one. More precisely, it does not refer to a closed curve, but rather to a conformation of the DNA chain, where two DNA sites that are widely separated from the genomic point of view are maintained close to each other by some other macromolecule, so that the segment of DNA located between these two sites looks like a loop when it is observed from a long distance or projected on a plane. However, the chain is not closed and torsional stress cannot diffuse directly from one extremity of the loop to the other one through the bridging macromolecule.

In order to ascertain this point, we launched a series of simulations with the crudest model for DNA-bridging proteins, namely a simple bond between beads $\alpha$ and $\beta$. The potential energy of the system was written as $E_{pot} + E_{BP}$, where $E_{pot}$ is the energy of the naked circular DNA chain described in Eqs. (S1)-(S5) and $E_{BP}$ an additional term, which models the action of DNA-bridging proteins. For the simplest model, we used

$$E_{BP} = \frac{h}{2}(d_{\alpha\beta} - d_{\alpha\beta}^0)^2 \ , \tag{3}$$

where $d_{\alpha\beta}$ is the distance between the centers of beads $\alpha$ and $\beta$ and $d_{\alpha\beta}^0 = 4$ nm. For the sake of simplicity, we used the same value of the stretching rigidity as for the DNA chain, that is $h = 100 \, k_B T / l_0^2$ (see Model and Simulation in the Supporting Material). The value $d_{\alpha\beta}^0 = 4$ nm is small enough to ensure that no other DNA segment crosses the segment between $\alpha$ and $\beta$, which is formally occupied by the DNA bridging protein. The energetic barrier for any segment to cross this line is indeed about $32.8 \, k_B T$, which is virtually insuperable. For the peace of mind, we nevertheless ran several simulations with $d_{\alpha\beta}^0 = 3$ nm, corresponding to a barrier of about $60.0 \, k_B T$, for the second model discussed below, and obtained exactly the same results as with $d_{\alpha\beta}^0 = 4$ nm. This confirms that Eq. (2) describes an impenetrable $\alpha\beta$ bond. The same initial conformations were used as in the preceding subsection and 100 different relaxation trajectories were again integrated for 1 ms for each initial conformation. Characteristic equilibration times obtained as described above are shown in Fig. S4. Comparison of Fig. S4 with Figs. 2 and S3 indicates that addition of a bond between beads $\alpha$ and $\beta$ affects significantly neither $\tau_{Tw}$ nor $\tau_{Wr}$. Stated in other words, *not all DNA-bridging proteins are topological barriers* or, equivalently, *not all DNA loops are independent topological domains*. The ability of certain DNA-bridging proteins, like LacI, to act as topological barriers is consequently related to some property, which is subtler than the mere connection between two genetically widely separated DNA sites.

Still, it is worth noting that the $\alpha\beta$ bond has a significant effect on the time evolution of $Wr_{12}$, as can be checked in Fig. S5. The reason is the following. In order to compute the partial writhes $Wr_j$ (see the Supporting Material), one has to introduce two "virtual" closed chains $C_1$ and $C_2$. $C_1$ is composed of beads $\alpha$ to $\beta - 1$, plus the "unphysical" segment between beads $\beta$ and $\alpha$, while $C_2$ is composed of beads $\beta$ to $\alpha - 1$, plus the same "unphysical" segment between beads $\alpha$ and $\beta$. When beads $\alpha$ and $\beta$ are bonded, they remain close to each other for $t > 0$, as indicated by the positions of the two arrows in Fig. 3(a), so that the unphysical segment between beads $\alpha$ and $\beta$ remains small and $C_1$ is a good representation of the initially torsionally relaxed DNA loop (and $Wr_1$ is a measure of its writhe), $C_2$ is a good representation of the initially torsionally stressed DNA loop (and $Wr_2$ is a measure of its writhe), and $Wr_{12}$ is a measure of the intrication of the two loops. In this case, $Wr_{12}$ remains close to zero at all times, as can be checked in Fig. S5(b). This indicates that, for bridged DNA, equilibration of the torsional stress is not accompanied by any significant intrication of the two circular moieties. In contrast, for the system without the $E_{BP}$ term, beads $\alpha$ and $\beta$ do not remain close to each other for $t > 0$, as indicated by the positions of the two arrows in Fig. 1(c), so that the unphysical segment between beads $\alpha$ and $\beta$ becomes arbitrarily large and $C_1$, $C_2$, the $Wr_j$ and $Wr_{12}$ lose their physical meaning. In this case $Wr_{12}$ varies quite widely with changing DNA conformations, as is illustrated in Fig. S5(a) for an initial conformation with $\Delta Lk^0 = -52$ ($\sigma^0 \approx -0.121$).

**Blocking diffusion of the twist is NOT sufficient to create independent topological domains**

The reason why forming a loop is not sufficient to create independent topological domains is of course that the bond between beads $\alpha$ and $\beta$ described in the previous sub-section is not able to block the rapid diffusion of the excess of twist through beads $\alpha$ and $\beta$ and the subsequent equilibration of the writhe. In order to form independent topological domains, the diffusion of the excess of twist must absolutely be blocked. Stated in other words, rotation of the internal basis $(\mathbf{u}_k, \mathbf{f}_k, \mathbf{v}_k)$ around $\mathbf{u}_k$ must be forbidden at $k = \alpha, \beta$. From the biological point of view, diffusion of the excess of twist is probably blocked because the involved proteins bind to their recognition site by inserting loops in the major or minor grooves of the DNA duplex. For example, it has been shown that H-NS dimers insert one C-terminal loop inside the minor groove of double-stranded DNA (59), thereby hindering rotation of the DNA around its own axis at this location. From the modeling point of view, diffusion of the twist can be blocked by imposing that $\mathbf{f}_\alpha$ and $\mathbf{f}_\beta$ remain perpendicular to $\mathbf{r}_{\alpha,\beta} = \mathbf{r}_\beta - \mathbf{r}_\alpha$ at all times ($\mathbf{f}_\alpha . \mathbf{r}_{\alpha,\beta} = \mathbf{f}_\beta . \mathbf{r}_{\alpha,\beta} = 0$), which is achieved by computing the angles

$$\delta_k = -\tan^{-1}(\frac{\mathbf{f}_k . \mathbf{r}_{\alpha,\beta}}{\mathbf{v}_k . \mathbf{r}_{\alpha,\beta}}) \, , \tag{4}$$

($k = \alpha, \beta$), after each integration step and rotating $\mathbf{f}_k$ and $\mathbf{v}_k$ around $\mathbf{u}_k$ by this angle $\delta_k$

$$\begin{aligned} \mathbf{f}_k &\to \cos\delta_k \mathbf{f}_k + \sin\delta_k \mathbf{v}_k \\ \mathbf{v}_k &\to -\sin\delta_k \mathbf{f}_k + \cos\delta_k \mathbf{v}_k \end{aligned} \qquad (5)$$

for $k = \alpha, \beta$. We note that the $\delta_k$ remain small, and the corrections in Eq. (5) consequently remain also small, only if $\mathbf{r}_{\alpha,\beta}$ is not perpendicular simultaneously to $\mathbf{f}_k$ and $\mathbf{v}_k$, that is if vectors $\mathbf{r}_{\alpha,\beta}$ and $\mathbf{u}_k$ ($k = \alpha, \beta$) are not collinear. In order to prevent such collinearity, two bending terms were added to $E_{BP}$, namely

$$E_{BP} = \frac{h}{2}(d_{\alpha\beta} - d_{\alpha\beta}^0)^2 + \frac{5g}{2}(\xi_\alpha - \frac{\pi}{2})^2 + \frac{5g}{2}(\xi_\beta - \frac{\pi}{2})^2 \,, \qquad (6)$$

where $\xi_\alpha$ denotes the angle formed by beads $\beta$, $\alpha$ and $\alpha+1$, and $\xi_\beta$ the angle formed by beads $\alpha$, $\beta$ and $\beta+1$. Note that the bending rigidity for these two angles was assumed to be five time larger than the bending rigidity of the DNA chain, in order for $\xi_\alpha$ and $\xi_\beta$ to deviate only moderately from $\pi/2$. In this second model, the action of DNA-bridging proteins is consequently modeled by the potential energy terms in Eq. (6), plus the small, *a posteriori* corrections in Eq. (5).

The relaxation dynamics for this second model is much slower than for the first one, so that each trajectory had to be integrated for as long as 30 ms. Four different relaxation trajectories were integrated for each initial conformation, and four different initial conformations were again used for each value of $\Delta Lk^0$. The typical time evolution of the twist and the writhe for an initial conformation with $\Delta Lk^0 = -20$ ($\sigma^0 \approx -0.047$) is shown in Fig. 4. At the level of single trajectories, the $Wr_j, Wr_{12}$ and the $\Delta Lk_j$ actually relax through a succession of plateaus followed by abrupt jumps, as is illustrated in Fig. S6 for $Wr_{12}$, while the time evolution is of course smoother when these quantities are averaged over several trajectories, as in Fig. 4(b). It is stressed that this behavior has no precise physical meaning but arises instead from the mathematical definition of the total writhe $Wr$ in Eq. (S11) and its somewhat arbitrary partitioning into the sum of the two partial quantities $Wr_j$ and $Wr_{12}$ in Eq. (S17). In particular, $Wr$ does not display any sharp jump along single trajectories.

As can be checked in Fig. 4, the second model of DNA-bridging proteins succeeds in preventing equilibration of the twist and the writhe in the two moieties of the chain. Indeed, $\Delta Tw_2 - \Delta Tw_1$ and $Wr_2 - Wr_1$ remain nearly constant (and $\Delta Lk_2 - \Delta Lk_1$ remains rigorously constant) throughout the integration interval. Nonetheless, this model still fails to describe topological barriers, because $Wr_1$ and $Wr_2$ do evolve in parallel until the limit $Wr_1 + Wr_2 = 0$ is reached (see Fig. 4(b)). According to Eq. (1), this limit is equivalent to $Wr_{12} = Wr$. The limit $Wr_{12} = Wr$ indicates very strong intrication of the two moieties. Indeed, examination of the final conformations, like the one with $\Delta Lk^0 = -40$ ($\sigma^0 \approx -0.093$) shown in Fig. 3(b), reveals very compact structures, with the initially torsionally relaxed (green) moiety winding around the initially torsionally stressed (red) moiety. Clearly, the intrication of the two moieties is achieved by transforming part of the plectonemic supercoils of the red moiety into toroidal supercoils. Another general feature of the final conformations, is that the portion of the red moiety not involved in the winding with the green moiety has a compact geometry that probably results from the superposition of plectonemic and toroidal supercoils.

More insight into the dynamics of the relaxation is gained by plotting the time evolution of $2\langle Wr_{12}\rangle / \Delta Lk^0$, where $\Delta Lk^0 / 2$ is the final linking number difference of the circular chain. The result is shown in Fig. 5, where $Wr_{12}$ has been averaged over the 16 relaxation trajectories that have been computed for each value of $\Delta Lk^0$. The four curves with $|\Delta Lk^0| \leq 34$ ($|\sigma^0| \leq 0.079$) superimpose at short times, which indicates that the relaxation speed is proportional to the total amount of torsional stress. However, for longer times and/or larger values of the linking number difference, a slower regime sets in, which is probably due to the resistance opposed by plectonemes against the reorganization of the chain and the intrication of the two moieties. Finally, we note that after full relaxation the writhe contribution accounts for approximately 90% of the linking number difference, which is significantly larger than the 70% contribution in the absence of domains (see Fig. S1).

In conclusion, this second model of DNA-bridging proteins indicates that blocking the diffusion of the excess of twist at the two DNA loci bound by the proteins is sufficient to maintain a constant *difference* between the linking number difference, excess of twist, and writhe of the two moieties, but that it is *not* sufficient to insure true topological independence. Indeed, the DNA chain evolves towards very compact and intricated final conformations, where the topology of each moiety is different from the initial one. The relaxation limit $Wr_{12} = Wr$ (or equivalently $Wr_1 + Wr_2 = 0$) appears non trivial and intriguing to us. We suspect that the explanation involves subtle considerations that should deserve further work.

**Topological barriers must block diffusion of the twist AND the relative orientation of the DNA segments**

The question that remains to be answered is consequently: What additional constraint must the DNA-bridging protein exert on its binding sites in order to preserve not only the *difference* in torsional stress between the two moieties of the DNA chain, but rather the precise initial torsional stress in each moiety ? The feeling that emerges upon examination of conformations like the one shown in Fig. 3(b), is that the winding of the initially relaxed moiety around the initially stressed one probably requires the relative rotation of the two DNA segments around the $\alpha\beta$ bond, and that no winding would be possible if this rotation were blocked. In order to check this conjecture, a third bending term was added to $E_{\text{BP}}$, namely

$$E_{\text{BP}} = \frac{h}{2}(d_{\alpha\beta} - d_{\alpha\beta}^0)^2 + \frac{5g}{2}(\xi_\alpha - \frac{\pi}{2})^2 + \frac{5g}{2}(\xi_\beta - \frac{\pi}{2})^2 + \frac{5g}{2}(\psi_{\alpha,\beta} - \psi_{\alpha,\beta}^0)^2 \;, \tag{7}$$

where $\psi_{\alpha,\beta}$ denotes the angle between vectors $\mathbf{r}_{\alpha,\alpha+1} = \mathbf{r}_{\alpha+1} - \mathbf{r}_\alpha$ and $\mathbf{r}_{\beta,\beta+1} = \mathbf{r}_{\beta+1} - \mathbf{r}_\beta$, and $\psi_{\alpha,\beta}^0$ the value of $\psi_{\alpha,\beta}$ at time $t = 0$. The new term in the right-hand side of Eq. (7) prevents large fluctuations of the relative orientations of the two DNA segments that are bridged by the protein. In particular, it blocks the rotation around the $\alpha\beta$ bond of one segment with respect to the other one. In this third model, the action of DNA-bridging proteins is consequently modeled by the potential energy terms in Eq. (7), plus the small, *a posteriori* corrections in Eq. (5).

The relaxation dynamics for this third model was investigated along the same lines as the second model. The typical time evolution of the twist and the writhe for an initial conformation with $\Delta Lk^0 = -52$ ($\sigma^0 \approx -0.121$) is shown in Fig. S7. As can be checked in this figure, all quantities, linking numbers difference, excess of twist, and writhe, whether local or global, remain strictly constant along the trajectories. No relaxation actually ever occurs and conformations obtained after long integration times are still composed of one torsionally relaxed moiety and one torsionally stressed moiety, as can be checked in Fig. 3(c). This third model of DNA-bridging proteins finally describes a topological barrier. In order to act as topological barriers, DNA-bridging proteins must consequently exert rather strong constraints on their binding sites: Indeed, they must block the diffusion of the excess of twist through both binding sites and must additionally block the rotation of one segment relative to the other one. It is worth emphasizing that this last condition is probably not satisfied by all DNA-bridging proteins. For example, it is well known that H-NS dimers bind most often non-specifically to the DNA, which means that the two C-terminal DNA-binding domains (59) located at the end of flexible linkers (60) bind into the minor groove of the DNA (59) through generic electrostatic forces instead of a precise chemical bond. The consequence is that H-NS dimers can slide along the DNA molecule (28,29), but they are also expected to rotate with respect to the DNA axis. Conversely, the two DNA segments can rotate with respect to the protein axis. Proteins binding specifically to the DNA are more likely to satisfy the no-rotation condition, provided that the bonds are strong enough to resist the torque exerted by the DNA. A second mechanism for satisfying the no-rotation condition is that the proteins systematically form pairs of neighboring bridges. In this respect, we note with interest that the authors of the experimental study about the topological barrier formed by LacI tetramers reported that the barrier formed by a single LacI tetramer *is much less stable comparing with those containing multiple LacI-lac O1 nucleoprotein complexes* (15). The most straightforward explanation for this result is that two bridges in tandem are required to sustain the torque arising from the difference in torsional stress in the two DNA loops. Yet, single molecule experiments have demonstrated that a single LacI bridge can sustain the torque associated with physiological values of supercoiling, that is $\sigma \approx -0.06$ (16). In this context, we surmise that the slow relaxation observed in (15) for single DNA-bridges may correspond to the slow intrication obtained with our second model for DNA-bridging proteins, while the more stable topological separation observed in (15) for bridges in tandem corresponds to true barriers obtained with the third model for DNA-bridging proteins. We finally note that the intrication properties discussed here are also expected to contribute to the formation of topological domains that are, in effect, larger than the domain delineated by the binding sites of LacI as recently observed in (16).

**CONCLUSIONS**

DNA-bridging proteins form DNA loops by dynamically cross-linking DNA sites that are widely separated from the genomic point of view (17-22). They usually bind DNA non-specifically but with high affinity. DNA-bridging proteins have now been investigated for more than two decades, but it is only very recently that it has been demonstrated that a DNA-bridging protein like the LacI repressor is able to separate circular plasmids into two topologically independent loops (15,23,24). To the best of our knowledge, this is at present the only demonstrated case of a DNA-bridging protein acting as a topological barrier. In the present work, we aimed at going one step further and determining the conditions that DNA-bridging proteins must fulfill to act as topological barriers. We showed that these proteins must exert rather strong constraints on their binding sites: They must block the diffusion of the excess of twist through the two binding sites on the DNA molecule and must simultaneously prevent the rotation of one DNA segment relative to the other one. These criteria should be of great help when investigating the (existence, or lack of) topological barrier properties of other DNA-bridging proteins, like for example H-NS, which has been predicted to participate in the organization of the genome into topologically independent domains (6,61). Indeed, it is probable that not all DNA-bridging proteins satisfy the second condition. If it is not satisfied, then intricate conformations are obtained, with a preserved difference in the torsional stress between the two domains, but different values of the stress in each domain. A rationale for the intriguing limit $Wr_{12} = Wr$ (or equivalently $Wr_1 + Wr_2 = 0$) is however missing.

This work also brings some light on another mechanism that has been proposed to explain the formation of independent topological domains, namely two proteins wrapping the DNA in two different locations (Ref. (15), Fig. 6). There is experimental evidence that certain wrapping proteins, like GalR (15), the λ O protein (15), and Fis (62), are able to form topological barriers by blocking the diffusion of twist. However, we checked that the torsional stress equilibrates rapidly in the two moieties of the DNA chain when simulations are performed with $d_{\alpha\beta}^0 = 10$ nm instead of 4 nm, because equilibration of twist and writhe is mediated in this case by the DNA chain passing repeatedly between beads $\alpha$ and $\beta$. The fact that GalR and λ O proteins are able to separate the plasmid into two stable topological domains (15) therefore suggests that the two wrapping proteins are located very close to each other or are bound by some impenetrable molecular complex. Moreover, their relative orientation remains probably constant, because the intrication mechanism would otherwise set in. Our work therefore clearly supports the hypothesis of the authors, who state that they *cannot fully exclude the role of protein-protein interactions of λ O-DNA complexes and GalR-DNA complexes in the formation of the two distinct topological domains* (15).

Still, our work suggests that the conditions for topological separation may be met by molecular geometries that do not require that the boundaries of the topological domains be close to each other. For example, one can imagine a topological barrier formed of two different proteins, which both anchor the DNA molecule to the cell membrane. According to our work, the DNA segment located between the two anchoring proteins may be topologically independent, provided that each protein is able to block the diffusion of twist along the DNA molecule and that the proteins cannot rotate with respect to the membrane. If this latter condition is not satisfied, then some complex winding of the DNA around itself is likely to occur, like for DNA-bridging proteins.

Another model which is frequently proposed for the formation of topological barriers, namely actively transcribing RNAP (10-12), is more difficult to discuss in terms of the results described in this paper, because several time scales need be considered. Indeed, it is likely that the two parts of the DNA molecule located respectively upstream and downstream from the transcribing RNAP will tend to increase their intricacy, in order to reduce the torsional stress induced by the blocking of the diffusion of twist at the RNAP. However, the RNAP is itself moving forward along the DNA, so that the ability of transcribing RNAP to act as topological barriers probably depends on the relative speeds of the two mechanisms. More work is needed to clarify this point. Similarly, the case of DNA-bending nucleoid proteins, like HU (63), certainly deserves further investigations.

Finally, we note with interest that the intricate structure obtained with DNA-bridging proteins that block the diffusion of twist along the DNA but do not prevent the relative rotation of the DNA segments are highly compact. It turns out that the mechanism leading to the formation of the bacterial nucleoid is a longstanding but still lively debated question (31,32,64-67). The point that has puzzled scientists for decades is that the volume of the unconstrained genomic DNA of bacteria in physiological solution, as estimated for example from the worm- like chain model (68), is approximately thousand times larger than the volume of the cell, while the nucleoid often occupies only a small fraction of the cell (69). It would certainly be interesting to check whether the intrication resulting from the alternation of torsionally stressed and torsionally relaxed DNA domains could contribute significantly to the compaction of the nucleoid.

**AUTHOR CONTRIBUTIONS**

I.J. imagined the research project and performed preliminary simulations. M.J. developed the models, ran the simulations and analyzed the results. M.J. and I.J. wrote the manuscript.

**ACKNOWLEDGMENTS**

M. J. thanks Prof. C. Hegedus (Université Lumière Lyon 2) for her hospitality during the Covid-19 epidemic confinement. I. J. thanks B. Houchmandzadeh (CNRS and Université Grenoble Alpes) for valuable discussions on the preliminary simulations.

**SUPPORTING MATERIAL**

Model and Simulations section. Figures S1 to S7.

**SUPPORTING CITATIONS**

References (70,71) appear in the Supporting Material.

# FIGURE CAPTIONS

**Figure 1** : **(a)** Representative conformation of an equilibrated circular chain with linking number difference $\Delta Lk^0 = -40$ ($\sigma^0 \approx -0.093$). The chain is separated formally into two moieties of equal length, which are colored in green and red, respectively. The black arrows locate beads $\alpha$ and $\beta$, which separate the two moieties. **(b)** Representative snapshot of the same chain after increasing the twist in the green segment by $\Delta Lk^0/2$ and letting this segment equilibrate again for 5 ms while keeping the red segment frozen. This is an out-of-equilibrium conformation. **(c)** Representative snapshot obtained upon relaxation of the out-of-equilibrium conformation shown in (b). The linking number difference of the equilibrated chain is $\Delta Lk^0/2 = -20$ ($\sigma^0 \approx -0.047$).

**Figure 2** : Evolution, as a function of $-\sigma^0$, of the characteristic time constants for the equilibration of **(a)** twist, $\tau_{Tw}$, and **(b)** writhe, $\tau_{Wr}$. Each point was obtained from a fit of Eq. (2) against the time evolution of $\Delta Tw_2 - \Delta Tw_1$ and $Wr_2 - Wr_1$ averaged over 100 trajectories starting from a single initial conformation (see Fig. S2). The four points shown for each value of $-\sigma^0$ correspond to four different initial conformations. The dot-dashed lines are linear fits to the data and are provided as guidelines to the eyes and as a means of comparison with Figs. S3 and S4.

**Figure 3** : **(a)** Representative snapshot obtained upon relaxation of the out-of equilibrium conformation shown in Fig. 1(b) under the constraint that beads $\alpha$ and $\beta$ are bonded (Eq. (3)). **(b)** Representative snapshot obtained upon relaxation of the out-of equilibrium conformation shown in Fig. 1(b) for the second model of DNA-bridging proteins (Eqs. (5) and (6)). **(c)** Representative snapshot obtained upon relaxation of the out-of equilibrium conformation shown in Fig. 1(b) for the third model of DNA-bridging proteins (Eqs. (5) and (7)). The linking number difference of the three conformations is $\Delta Lk^0/2 = -20$.

**Figure 4** : Time evolution of **(a)** the excess of twist, and **(b)** the writhe, during the relaxation of a torsionally out-of-equilibrium DNA conformation with $\Delta Lk^0 = -20$ ($\sigma^0 \approx -0.047$) constrained by the second model for DNA-bridging proteins (Eqs. (5) and (6)). Each curve was averaged over 4 different trajectories starting from the same initial conformation.

**Figure 5 :** Time evolution of $2\langle Wr_{12}\rangle / \Delta Lk^0$ during the relaxation of torsionally out-of-equilibrium DNA chains constrained by the second model for DNA-bridging proteins (Eqs. (5) and (6)). Each curve was averaged over the 16 different trajectories with the same value of $\Delta Lk^0$. The value of $\Delta Lk^0$ is indicated on each curve.

**FIGURE 1**

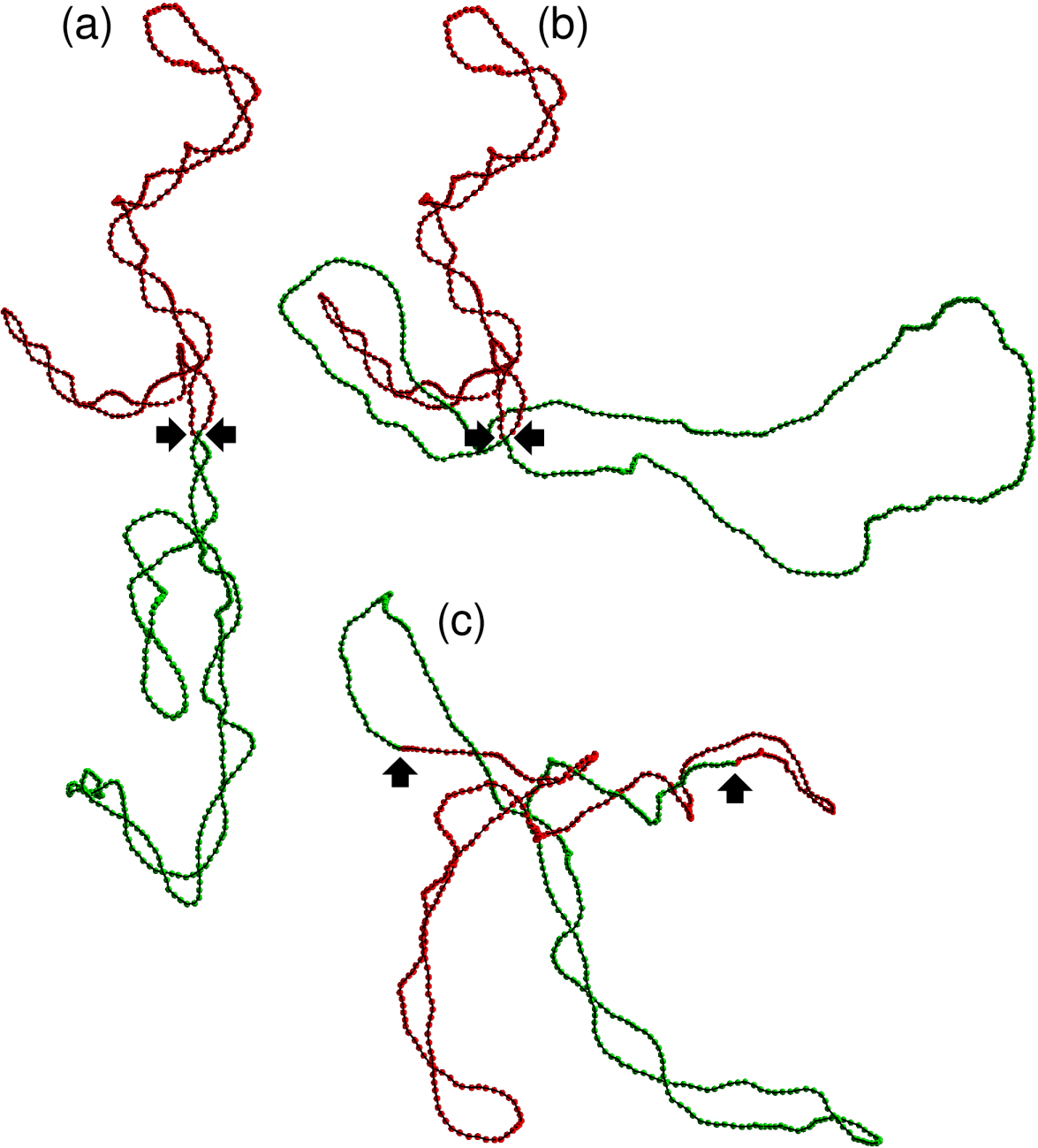

**FIGURE 2**

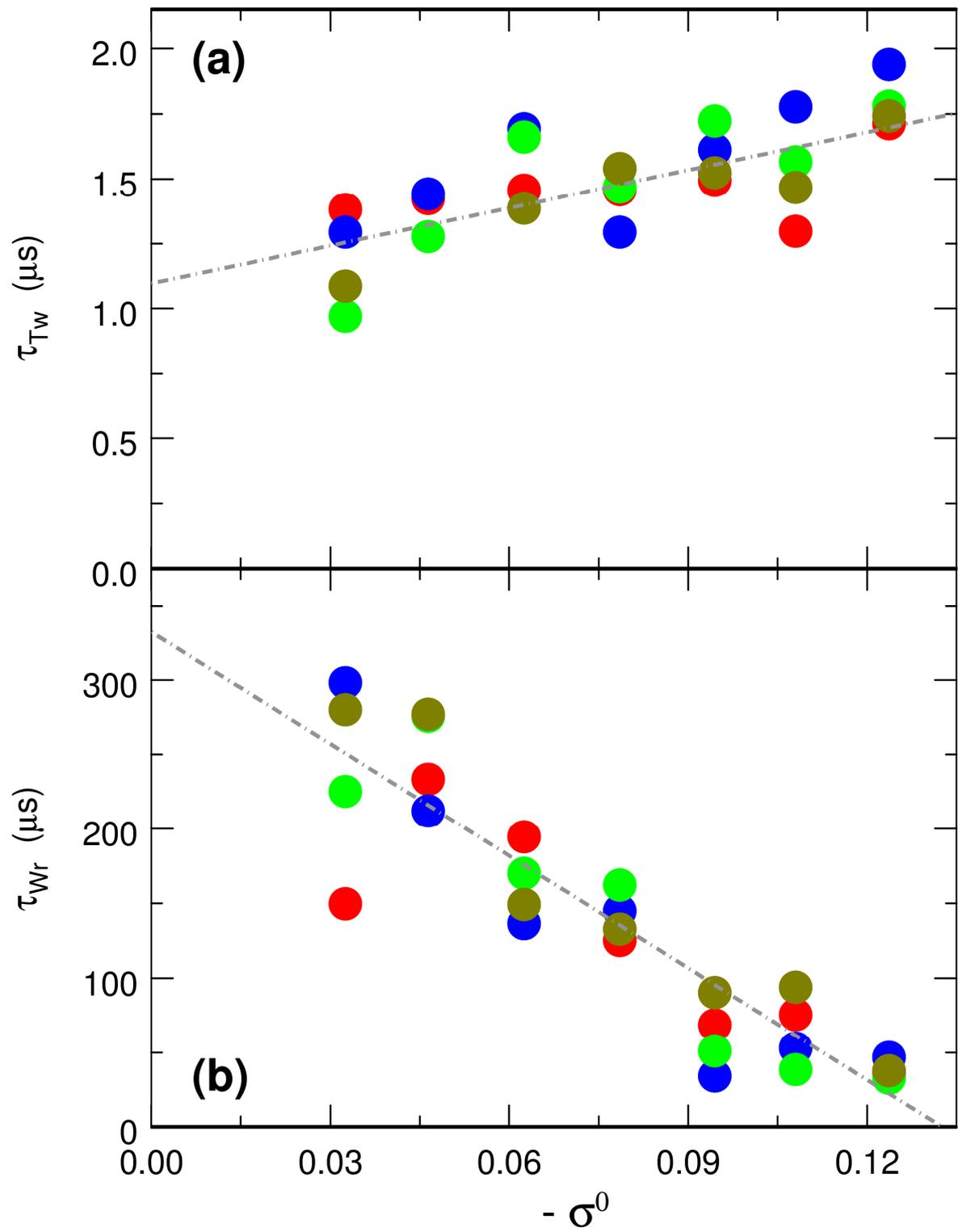

**FIGURE 3**

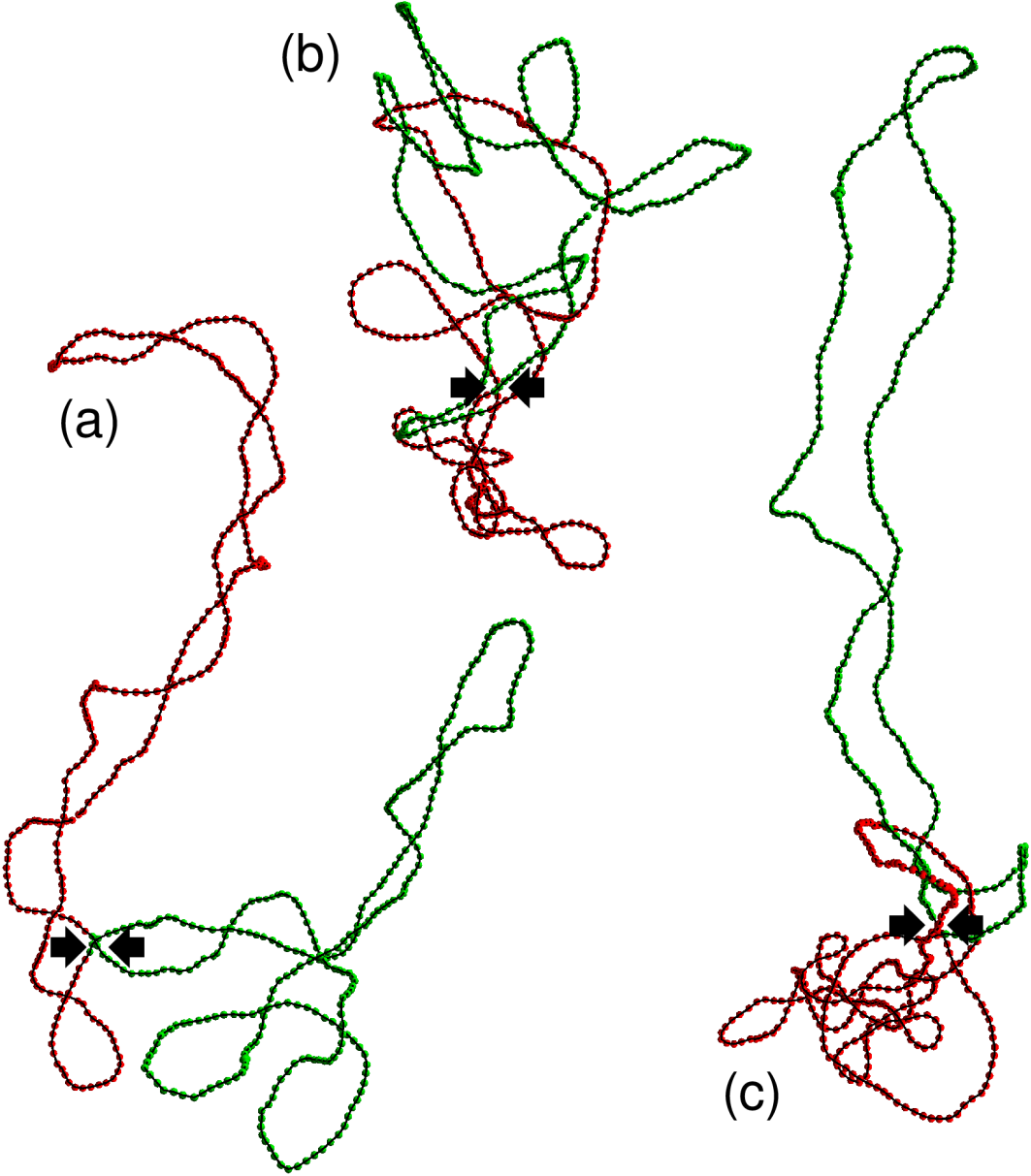

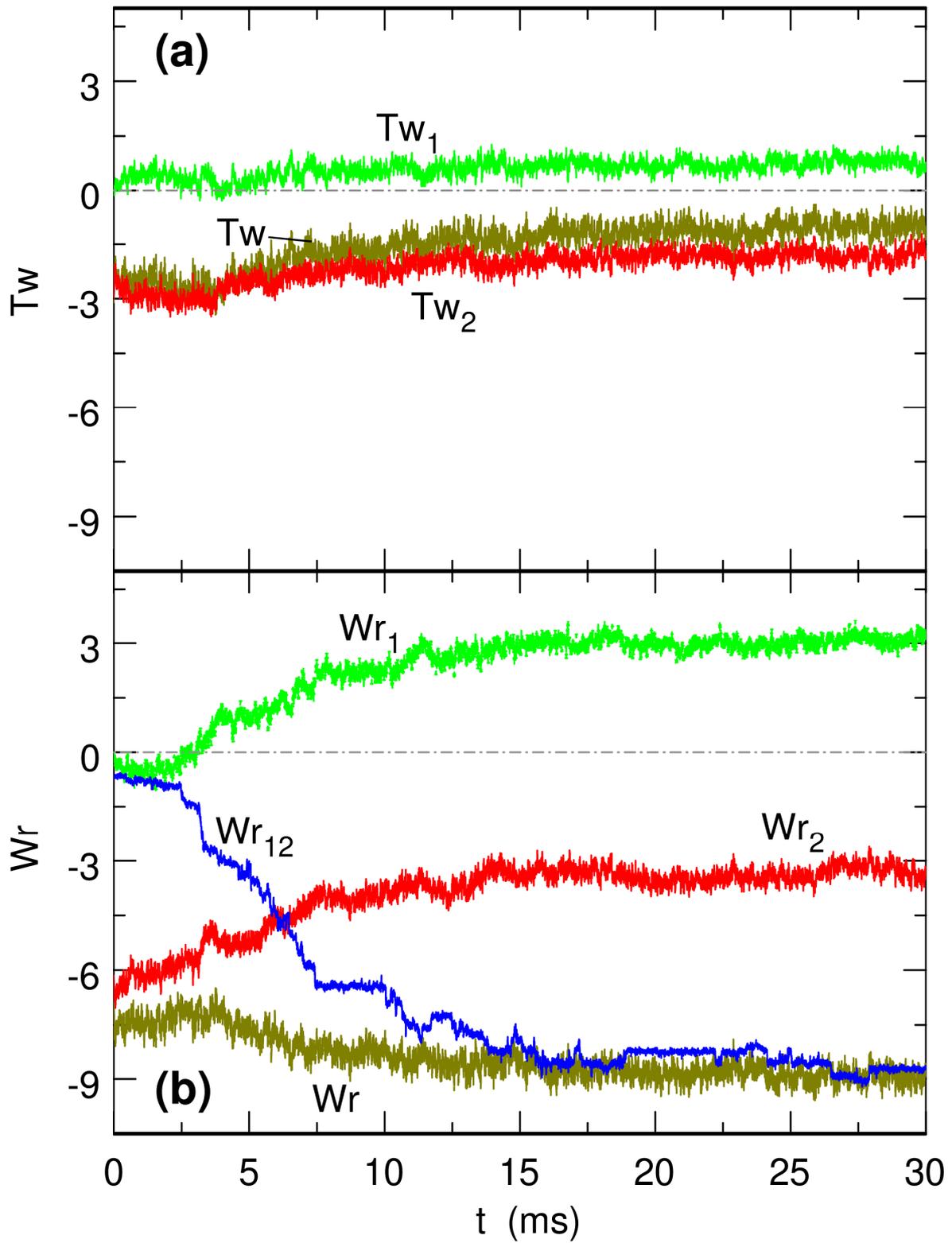

**FIGURE 5**

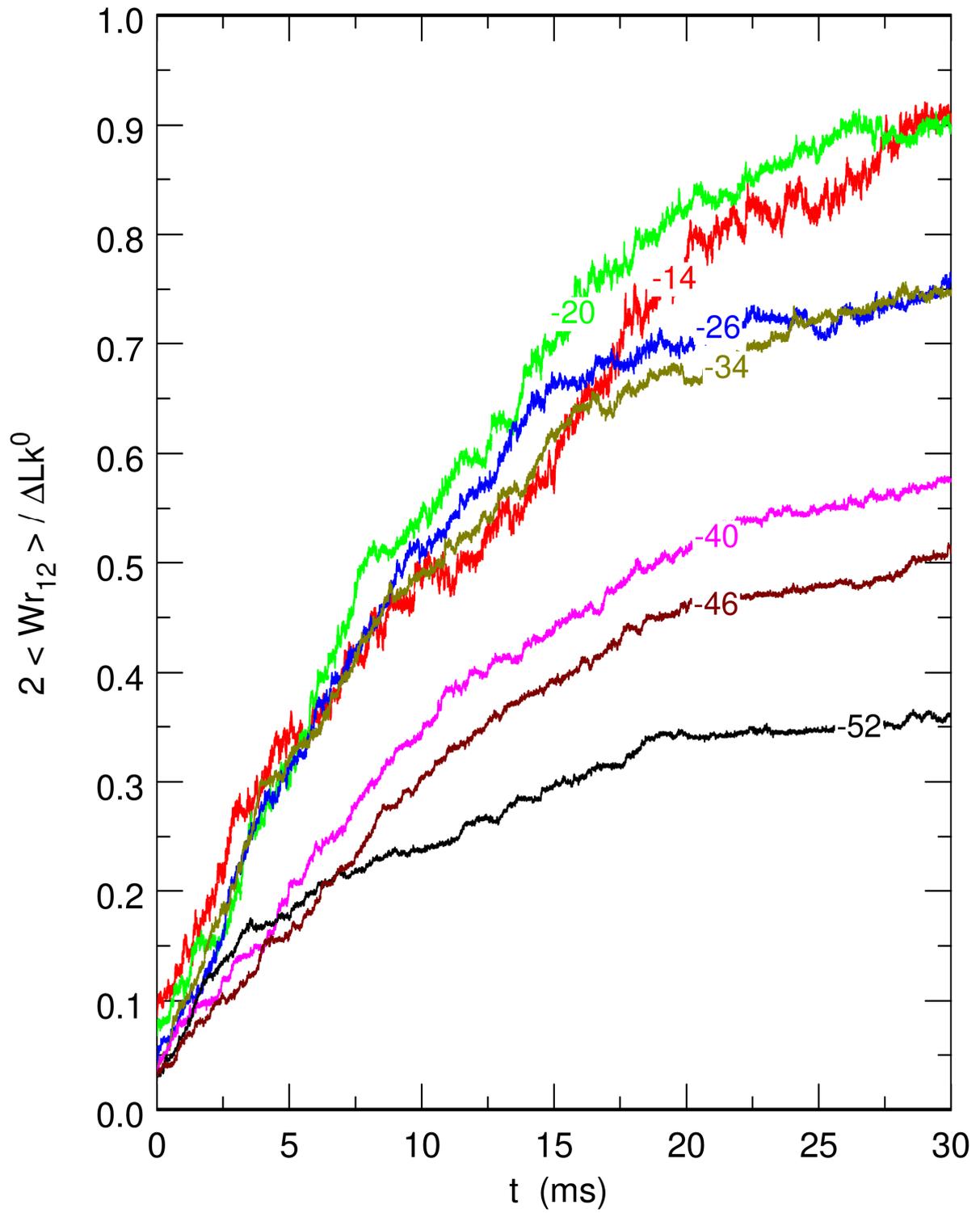

# Requirements for DNA-bridging proteins to act as topological barriers of the bacterial genome

- Supporting Material –

Marc Joyeux

*Laboratoire Interdisciplinaire de Physique,*

*CNRS and Université Grenoble Alpes, Grenoble, France*

Ivan Junier

*TIMC-IMAG*

*CNRS and Université Grenoble Alpes, Grenoble, France*

## MODEL AND SIMULATIONS

Temperature $T$ is assumed to be 298 K throughout the study. Circular DNA molecules are modeled as circular chains of $n=600$ beads with radius $a=1.0$ nm separated at equilibrium by a distance $l_0=2.5$ nm. Two beads represent 15 DNA base pairs (bp) and chains of $n=600$ beads (equivalent to 4500 bp) model the plasmids investigated in (1). The potential energy of the system, $E_{\text{pot}}$, consists of four terms

$$E_{\text{pot}} = V_s + V_b + V_t + V_e \ , \tag{S1}$$

where

$$V_s = \frac{h}{2}\sum_{k=1}^{n}(l_k - l_0)^2 \tag{S2}$$

$$V_b = \frac{g}{2}\sum_{k=1}^{n}\theta_k^2 \tag{S3}$$

$$V_t = \frac{\tau}{2}\sum_{k=1}^{n}(\Phi_{k+1} - \Phi_k)^2 \tag{S4}$$

$$V_e = q^2 \sum_{k=1}^{n-2}\sum_{K=k+2}^{n} H(\|\mathbf{r}_k - \mathbf{r}_K\| - 2a) \tag{S5}$$

describe the stretching, bending, torsional, and electrostatic energy of the DNA chain, respectively. In Eqs. (S2)-(S5), $\mathbf{r}_k$ denotes the position of bead $k$ (with the convention that $\mathbf{r}_{n+k} = \mathbf{r}_k$), $l_k = \|\mathbf{r}_{k+1} - \mathbf{r}_k\|$ the distance between two successive beads $k$ and $k+1$, and $\theta_k = \arccos((\mathbf{r}_{k+1} - \mathbf{r}_k)(\mathbf{r}_{k+2} - \mathbf{r}_{k+1}) / (\|\mathbf{r}_{k+1} - \mathbf{r}_k\| \|\mathbf{r}_{k+2} - \mathbf{r}_{k+1}\|))$ the angle formed by three successive beads $k$, $k+1$, and $k+2$.

The stretching energy $V_s$ is a computational device without biological meaning, which is aimed at avoiding a rigid rod description. The stretching rigidity $h$ was set to $h = 100\, k_B T / l_0^2$ to insure that the variations of the distance between successive beads remain small enough (2).

In contrast, the bending rigidity $g$ in the expression of $V_b$ was obtained from the known persistence length of DNA, $\xi = 50$ nm, according to $g = \xi\, k_B T / l_0 = 20\, k_B T$.

The torsional contribution $V_t$ was borrowed from Ref. (3) and torsion forces and momenta were computed as described therein. $\Phi_{k+1} - \Phi_k$ denotes the rotation of the body-fixed frame $(\mathbf{u}_k, \mathbf{f}_k, \mathbf{v}_k)$ around $\mathbf{u}_k$ between successive beads $k$ and $k+1$, where $\mathbf{u}_k = (\mathbf{r}_{k+1} - \mathbf{r}_k) / \|\mathbf{r}_{k+1} - \mathbf{r}_k\|$ is the unit vector pointing from bead $k$ to bead $k+1$. By convention, $\Phi_{n+k} = \Phi_k$. The value of the torsional rigidity, $\tau = 25\, k_B T$, was obtained by imposing that the writhe contribution $Wr$ accounts for approximately 70% of the linking number difference $\Delta Lk$ at equilibrium (4), see Fig. S1. The values of the bending and torsional rigidities of the model are close to each other, which agrees with experimental findings (4).

The electrostatic energy of the DNA chain, $V_e$, is expressed as a sum of repulsive Debye-Hückel terms with hard core. Function $H(r)$ is defined according to

$$H(r) = \frac{1}{4\pi\varepsilon r} \exp\left(-\frac{r}{r_D}\right) , \tag{S6}$$

where $\varepsilon = 80\, \varepsilon_0$ denotes the dielectric constant of the buffer and $r_D = 1.07$ nm the Debye length inside the buffer. This value of the Debye length corresponds to a concentration of monovalent salt of 100 mM, which is the value that is generally assumed for the cytoplasm of bacterial cells. $q$ is the value of the electric charge, which is placed at the centre of each DNA bead

$$q = -\frac{l_0 \overline{e}}{\ell_B} \approx -3.52 \, \overline{e}, \tag{S7}$$

where $\overline{e}$ is the absolute charge of the electron and $\ell_B = 0.7$ nm the Bjerrum length of water. In Eq. (S7), $\overline{e}/\ell_B$ is the net linear charge density along a DNA molecule immersed in a buffer with monovalent cations derived from Manning's counterion condensation theory (5,6). Note that electrostatic interactions between nearest neighbours are not included in Eq. (S5) because it is considered that they are already accounted for in the stretching and bending terms.

The dynamics of the system was investigated by integrating numerically overdamped Langevin equations. Practically, the updated positions and torsion angles at time step $i+1$ are computed from the positions and torsion angles at time step $i$ according to

$$\mathbf{r}_k^{(i+1)} = \mathbf{r}_k^{(i)} + \frac{\Delta t}{6\pi\eta a} \mathbf{f}_k^{(i)} + \sqrt{\frac{2 k_B T \Delta t}{6\pi\eta a}} \, x_k^{(i)} \tag{S8}$$

$$\Phi_k^{(i+1)} = \Phi_k^{(i)} + \frac{\tau \Delta t}{4\pi\eta a^2 l_0}(\Phi_{k+1}^{(i)} - 2\Phi_k^{(i)} + \Phi_{k-1}^{(i)}) + \sqrt{\frac{2 k_B T \Delta t}{4\pi\eta a^2 l_0}} \, X_k^{(i)}, \tag{S9}$$

where $\mathbf{f}_k^{(i)}$ are vectors of inter-particle forces arising from the potential energy $E_{pot}$, $T = 298$ K is the temperature of the system, $x_k^{(i)}$ and $X_k^{(i)}$ are vectors of random numbers extracted from a Gaussian distribution of mean 0 and variance 1, $\eta = 0.00089$ Pa s is the viscosity of the buffer at 298 K, and $\Delta t = 10$ ps is the integration time step.

The excess of twist $\Delta Tw$, the writhe $Wr$, and the linking number difference $\Delta Lk$ of the DNA chain were computed at regular time intervals according to (3)

$$\Delta Tw = \frac{1}{2\pi} \sum_{k=1}^{n} (\Phi_{k+1} - \Phi_k) \tag{S10}$$

$$Wr = \frac{1}{4\pi} \sum_{k=1}^{n} \sum_{\substack{K=1 \\ K \neq k}}^{n} \frac{[(\mathbf{r}_{k+1} - \mathbf{r}_k) \times (\mathbf{r}_{K+1} - \mathbf{r}_K)] \cdot (\mathbf{r}_k - \mathbf{r}_K)}{|\mathbf{r}_k - \mathbf{r}_K|^3} \tag{S11}$$

$$\Delta Lk = \Delta Tw + Wr. \tag{S12}$$

The superhelical density $\sigma$ was subsequently estimated from

$$\sigma = \frac{\Delta Lk}{Lk_0}, \tag{S13}$$

where the linking number $Lk_0 = 7.5 n / 10.5 \approx 429$ is the ratio of the number of base pairs of the DNA chain and the mean number of base pairs per turn of the torsionally relaxed double helix. Fig. S1 shows the plot of $\langle Wr / \Delta Lk \rangle$, the mean contribution of the writhe to the linking number difference, as a function of $-\sigma$. This plot is in good agreement both with experimental results (4) and results obtained with another model (7).

Partial values of the excess of twist and the writhe were also computed for each of the two moieties separated by beads $\alpha$ ($1 \leq \alpha \leq n/2$) and $\beta = \alpha + n/2$ (each moiety comprises $n/2$ beads). Partial values of the excess of twist were simply obtained from

$$\Delta Tw_1 = \frac{1}{2\pi} \sum_{k=\alpha}^{\beta-1} (\Phi_{k+1} - \Phi_k) \tag{S14}$$

$$\Delta Tw_2 = \frac{1}{2\pi} \sum_{k=\beta}^{\beta+n/2-1} (\Phi_{k+1} - \Phi_k) . \tag{S15}$$

The sum of $\Delta Tw_1$ and $\Delta Tw_2$ is obviously equal to the total excess of twist

$$\Delta Tw = \Delta Tw_1 + \Delta Tw_2 . \tag{S16}$$

The case of the writhe is more complex. Partial values were obtained by defining two closed chains of length $n/2$, namely $C_1 = (\mathbf{r}_\alpha, \mathbf{r}_{\alpha+1}, \mathbf{r}_{\alpha+2}, ..., \mathbf{r}_{\beta-2}, \mathbf{r}_{\beta-1}, \mathbf{r}_\alpha)$ and $C_2 = (\mathbf{r}_\beta, \mathbf{r}_{\beta+1}, \mathbf{r}_{\beta+2}, ..., \mathbf{r}_n, \mathbf{r}_1, ..., \mathbf{r}_{\alpha-2}, \mathbf{r}_{\alpha-1}, \mathbf{r}_\beta)$, and computing the writhe for $C_1$ and $C_2$ (labeled $Wr_1$ and $Wr_2$, respectively) according to equations similar to Eq. (S11). Note that $C_1$ and $C_2$ are "virtual" chains, because torsional stress cannot flow directly between beads $\alpha$ and $\beta$. Most importantly, $Wr$ is not necessarily close to the sum of $Wr_1$ and $Wr_2$, because of the double sums in the expressions for $Wr$, $Wr_1$, and $Wr_2$. Instead, one has

$$Wr = Wr_1 + Wr_2 + Wr_{12} . \tag{S17}$$

$Wr$, $Wr_1$, and $Wr_2$ quantify the winding of each closed chain around itself. In contrast, $Wr_{12}$ quantifies the winding of $C_1$ around $C_2$, that is, loosely speaking, the intrication of the two loops when beads $\alpha$ and $\beta$ remain close to each other.

Finally, the global torsional state of each moiety can be described by the value of the partial linking number difference

$$\Delta Lk_1 = \Delta Tw_1 + Wr_1 \tag{S18}$$

$$\Delta Lk_2 = \Delta Tw_2 + Wr_2 . \tag{S19}$$

It results from Eqs. (S12) and (S16)-(S19) that

$$\Delta Lk = \Delta Lk_1 + \Delta Lk_2 + Wr_{12} . \tag{S20}$$

# SUPPORTING REFERENCES

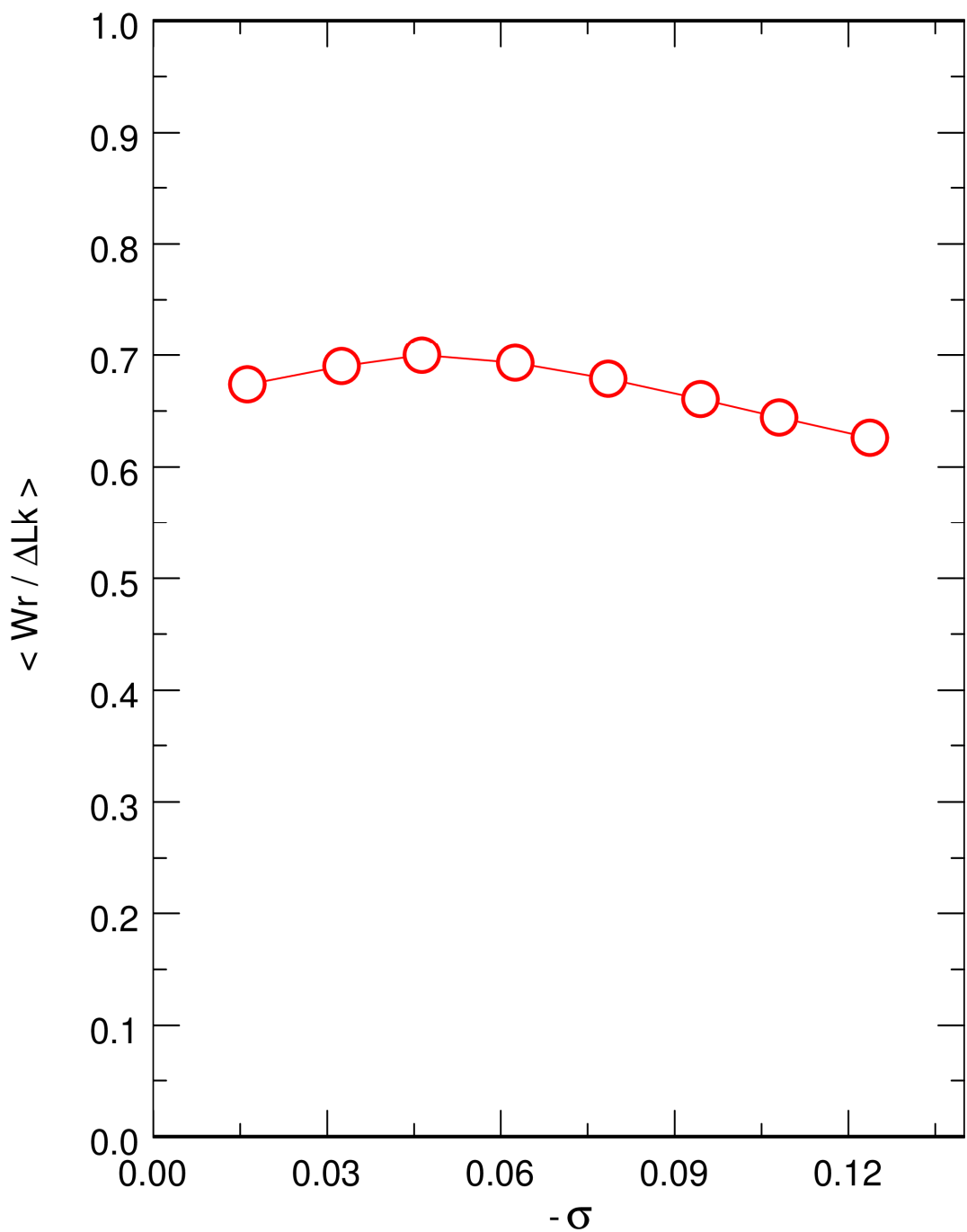

**Figure S1** : Evolution of $\langle Wr/\Delta Lk \rangle$, the mean contribution of the writhe to the linking number difference, as a function of $-\sigma$.

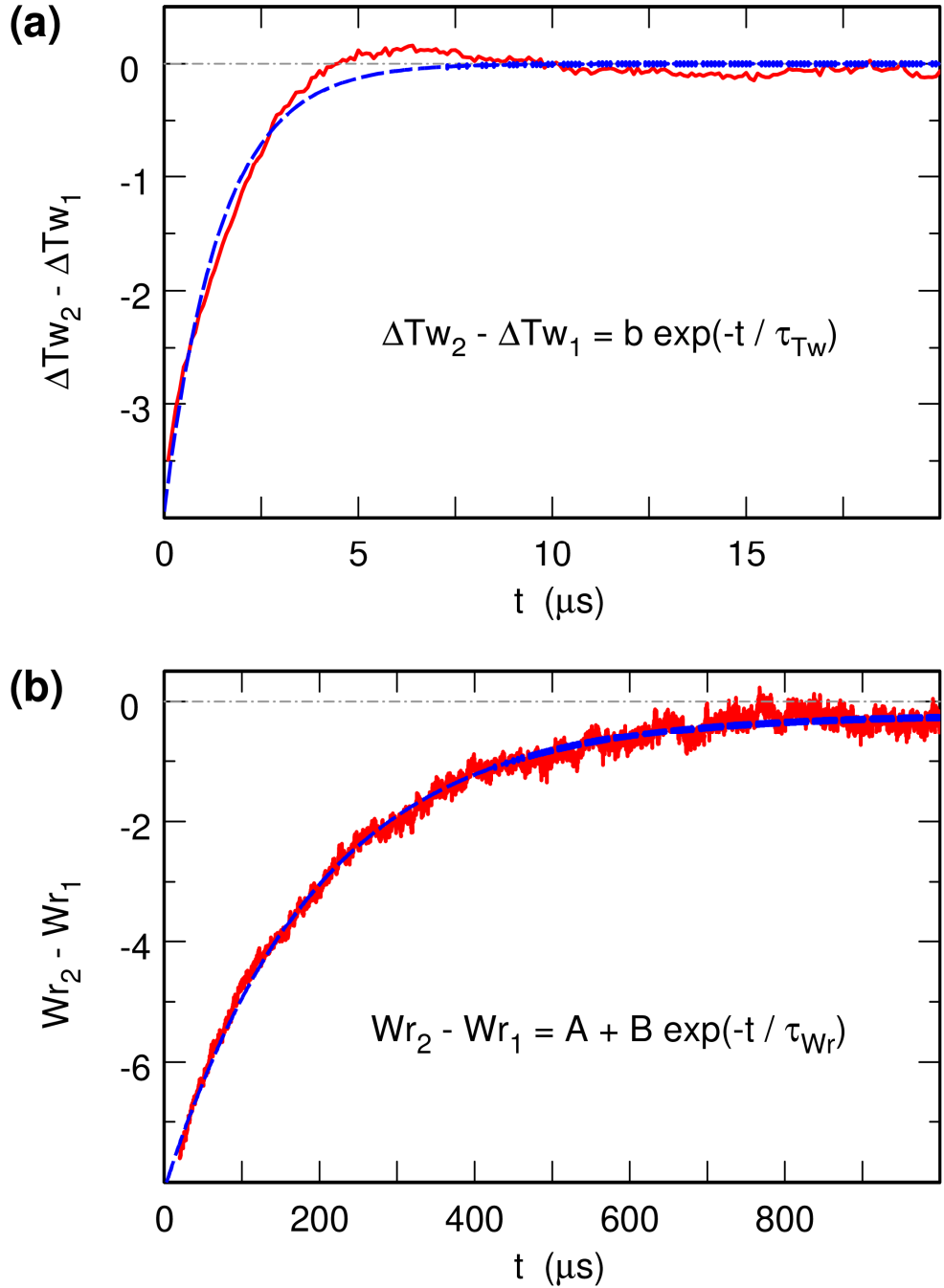

**Figure S2** : Time evolution of **(a)** $\Delta Tw_2 - \Delta Tw_1$, and **(b)** $Wr_2 - Wr_1$, for an initial conformation with $\Delta Lk^0 = -26$ ($\sigma^0 \approx -0.061$). The solid red lines show the results obtained upon averaging of $\Delta Tw_2 - \Delta Tw_1$ and $Wr_2 - Wr_1$ over 100 different trajectories starting from the same initial conformation. The dashed blue lines show the result of exponential fits against the red lines. The time windows used in the fits are $0 \le t \le 20$ μs for $\Delta Tw_2 - \Delta Tw_1$ and $10 \le t \le 1000$ μs for $Wr_2 - Wr_1$.

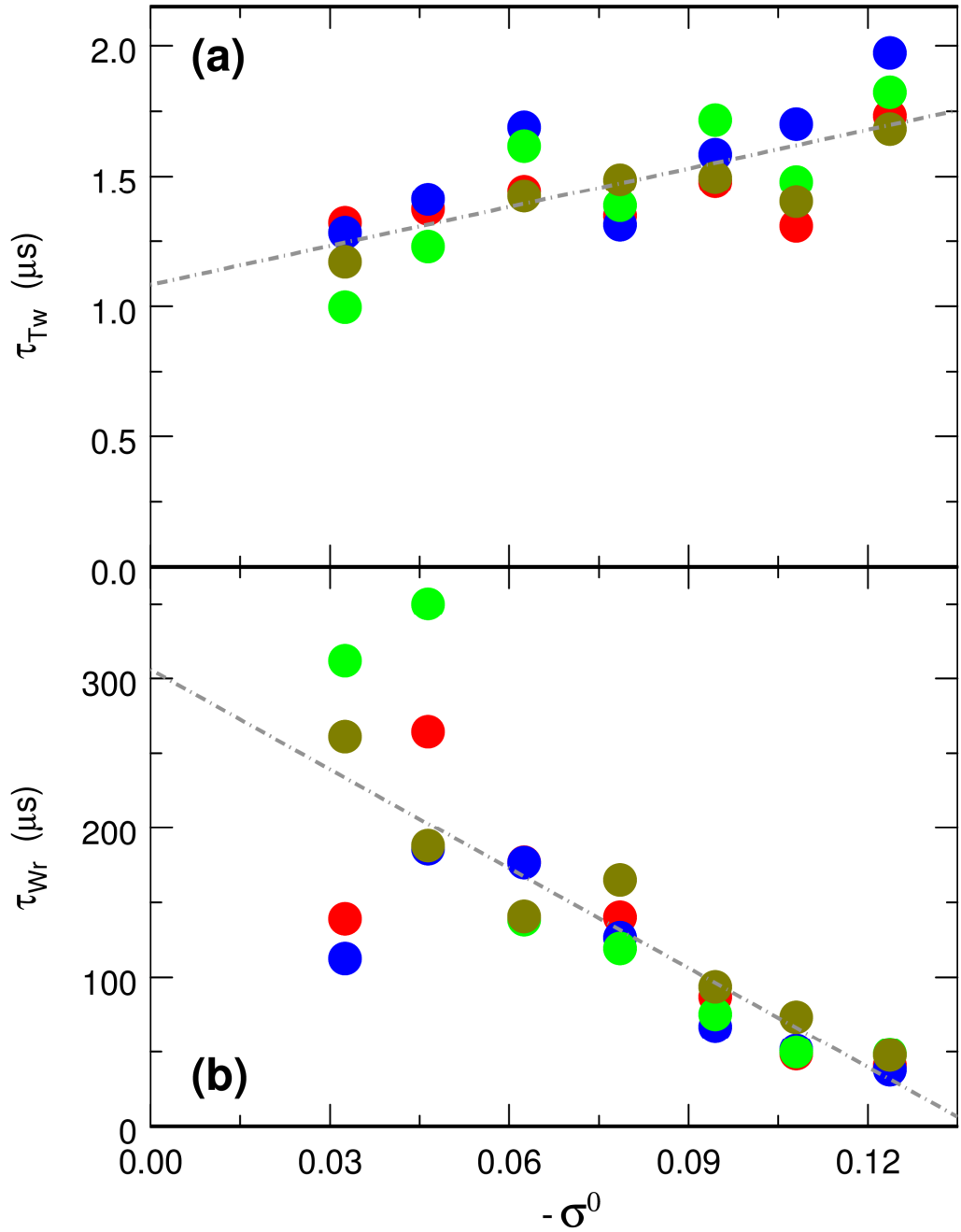

**Figure S3**: Same as Fig. 2, except that rotational noise has been discarded from evolution equations Eq. (S9). Plots **(a)** and **(b)** show the evolution of $\tau_{Tw}$ and $\tau_{Wr}$ as a function of $-\sigma^0$, respectively. The dot-dashed lines are linear fits to the data and are provided as guidelines to the eyes and as a means of comparison with Figs. 2 and S4.

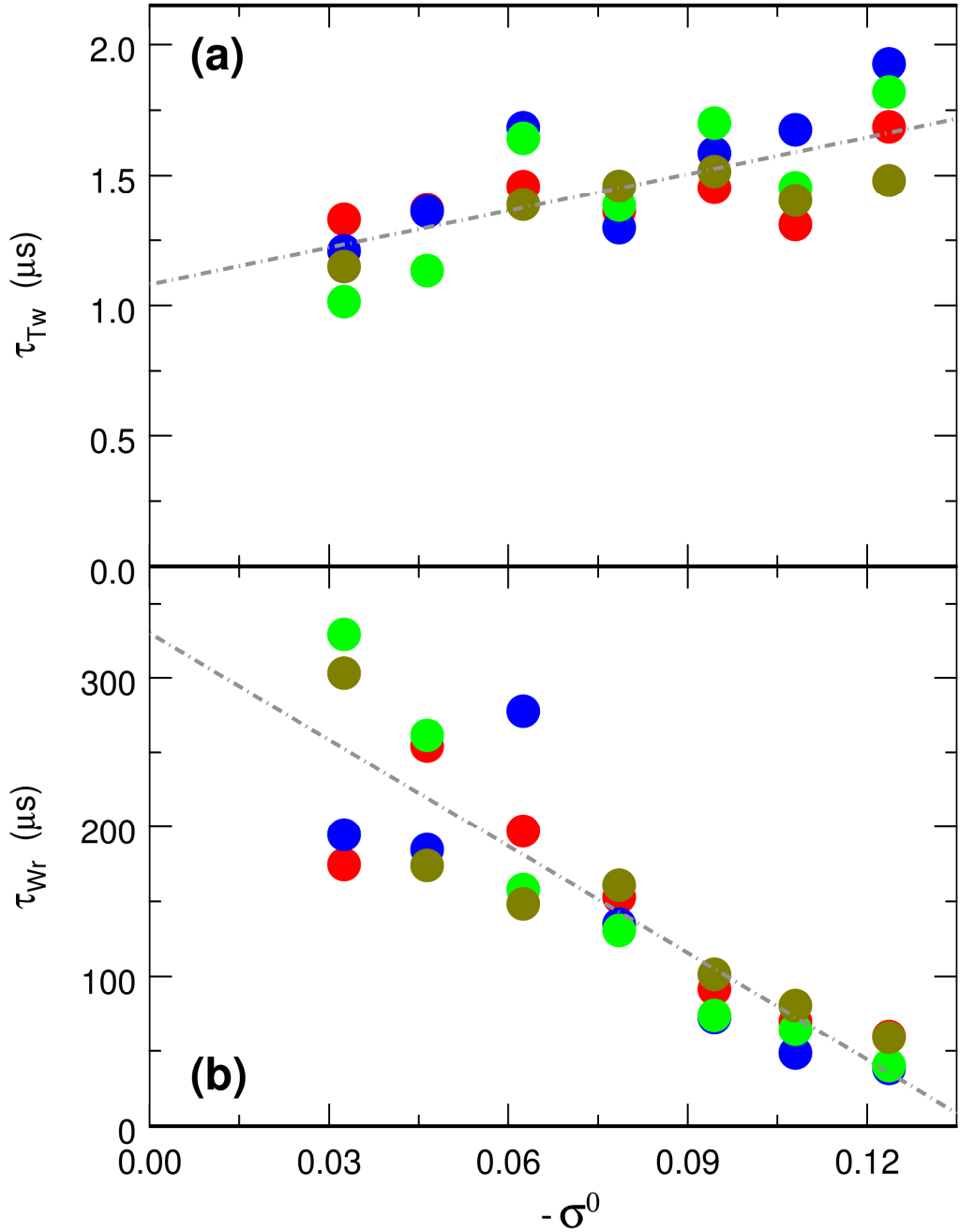

**Figure S4** : Same as Fig. S3, except that a bond has been added between beads $\alpha$ and $\beta$, as the crudest model for a DNA-bridging protein (see Eq. (3)). Plots **(a)** and **(b)** show the evolution of $\tau_{Tw}$ and $\tau_{Wr}$ as a function of $-\sigma^0$, respectively. The dot-dashed lines are linear fits to the data and are provided as guidelines to the eyes and as a means of comparison with Figs. 2 and S3.

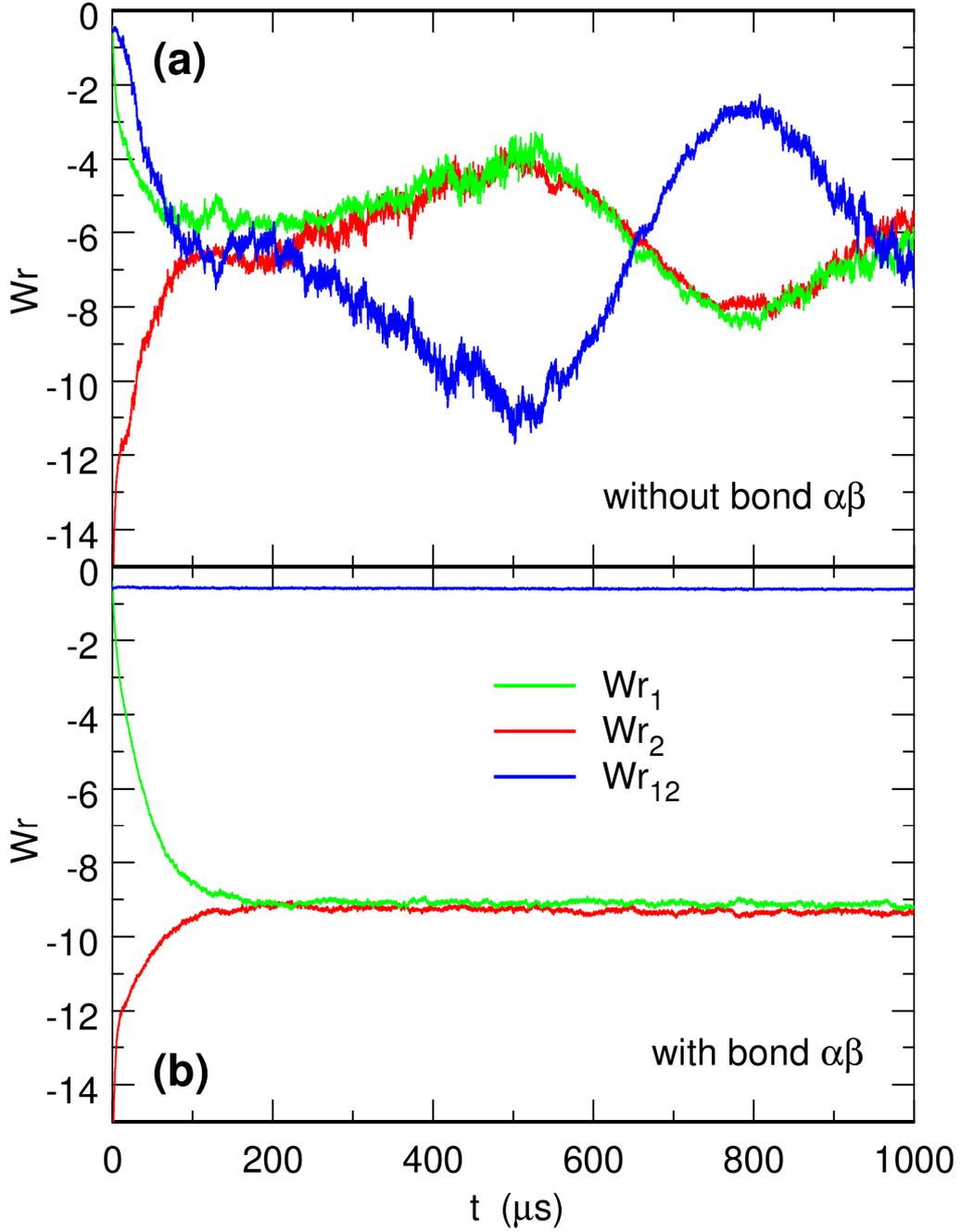

**Figure S5** : Time evolution of $Wr_1$, $Wr_2$, and $Wr_{12}$ during relaxation of a torsionally out-of-equilibrium DNA conformation with $\Delta Lk^0 = -52$ ($\sigma^0 \approx -0.121$) and **(a)** no constraint on the DNA chain, or **(b)** a bond between beads $\alpha$ and $\beta$ (Eq. (3)). Each plot was obtained by averaging $Wr_1$, $Wr_2$, and $Wr_{12}$ over 100 different trajectories starting from the same initial conformation.

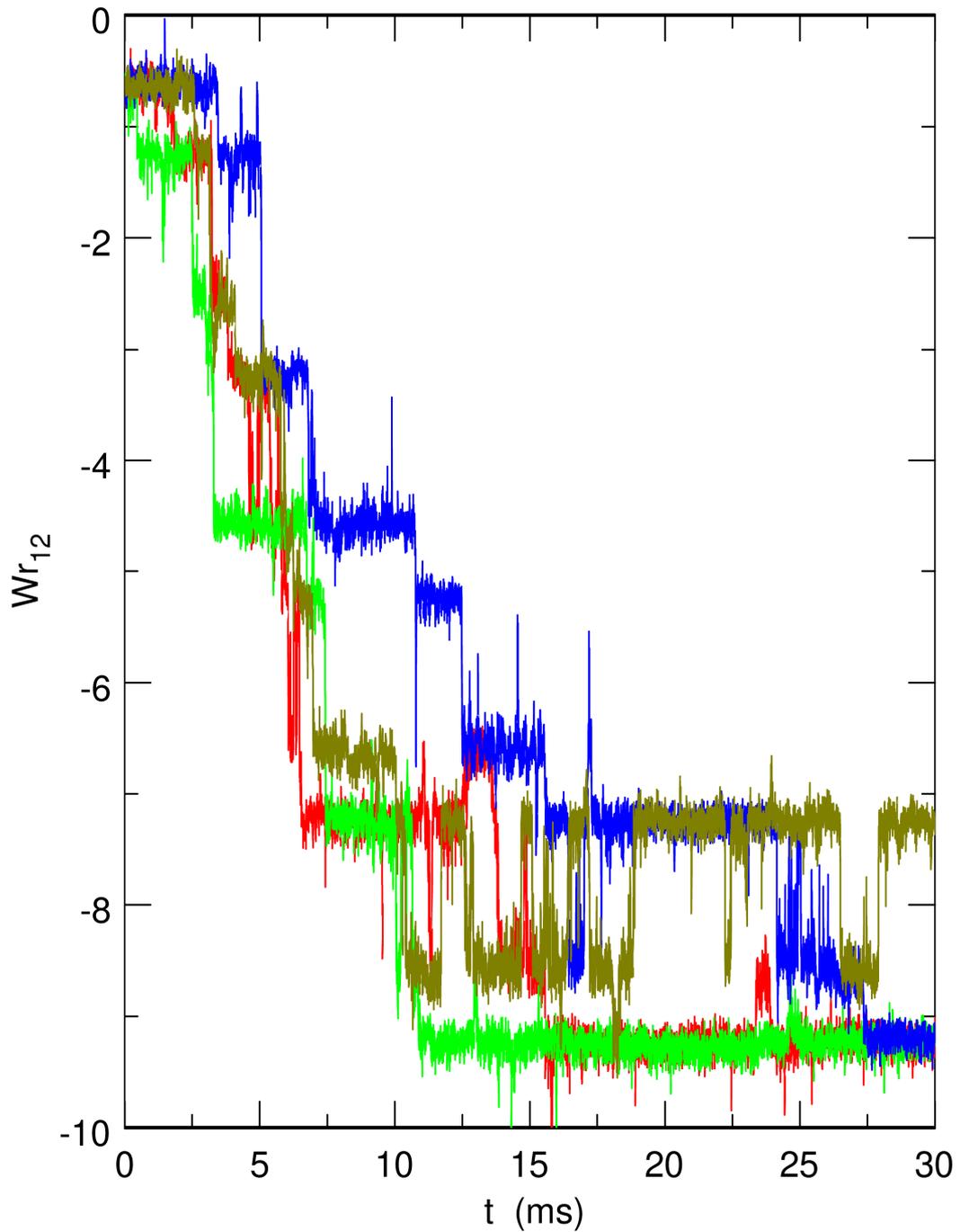

**Figure S6** : Time evolution of $Wr_{12}$ during the relaxation of a torsionally out-of-equilibrium DNA conformation with $\Delta Lk^0 = -20$ ($\sigma^0 \approx -0.047$) constrained by the second model for DNA-bridging proteins (Eqs. (5) and (6)). The four curves correspond to four different trajectories starting from the same initial conformation.

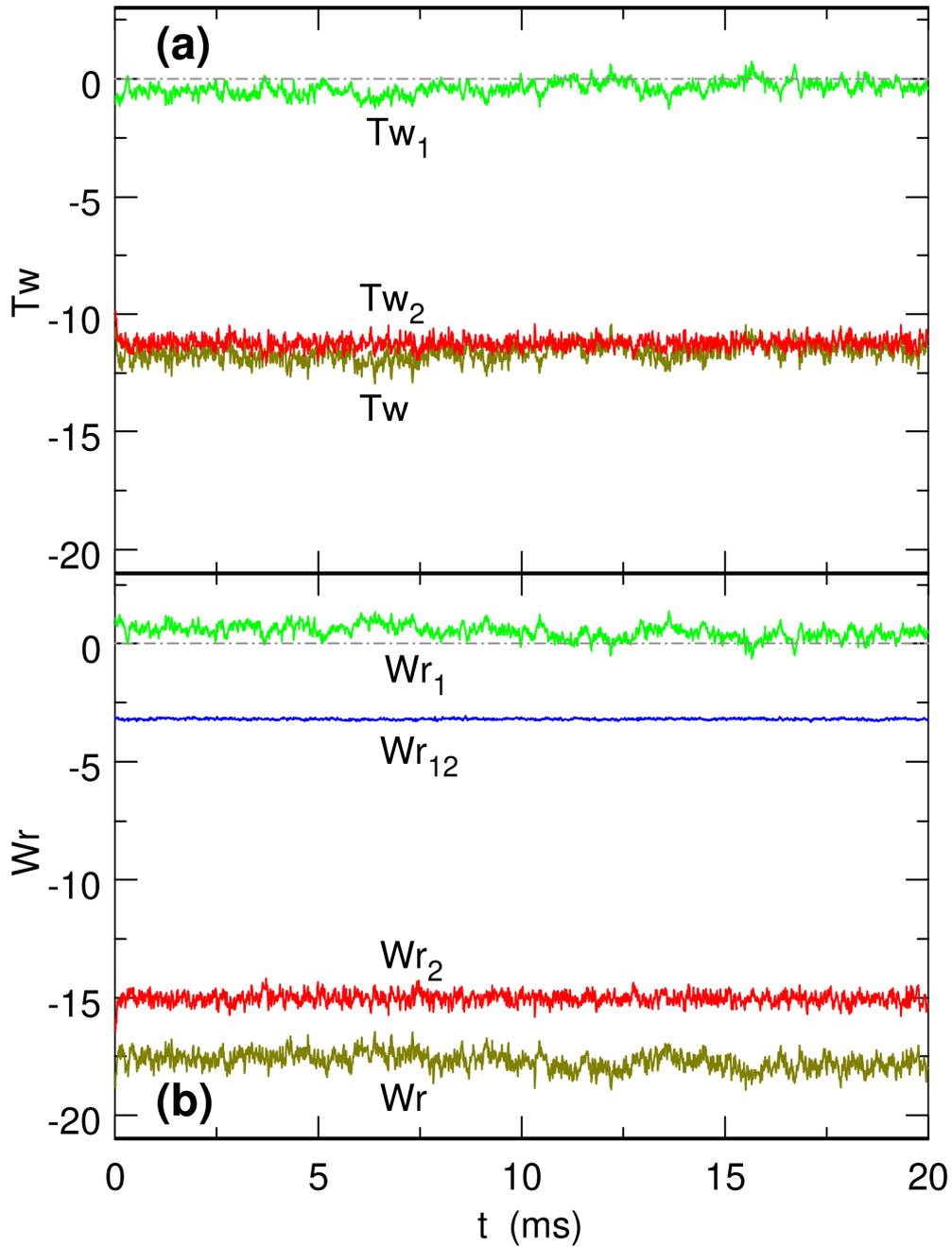

**Figure S7 :** Time evolution of **(a)** the excess of twist, and **(b)** the writhe, during the relaxation of a torsionally out-of-equilibrium DNA conformation with $\Delta Lk^0 = -52$ ($\sigma^0 \approx -0.121$) constrained by the third model for DNA-bridging proteins (Eqs. (5) and (7)). Each curve was averaged over 2 different trajectories starting from the same initial conformation.